# An Efficient Method for Optimizing RFID Reader Deployment and Energy Saving

Ching-Hsien Hsu[1*], Daqiang Zhang[2], Chao-Tung Yang[3] and Hai-Cheng Chu[4]

[1]Department of Computer Science and Information Engineering, Chung Hua University, Hsinchu, Taiwan
chh@chu.edu.tw
[2]School of Software Engineering, Tongji University, Shanghai, China
dqzhang@ieee.org
[3]Department of Computer Science, Tunghai University, Taichung, Taiwan
ctyang@thu.edu.tw
[4]Department of International Business, National Taichung University of Education, Taichung, Taiwan
ayura66@gmail.com

ABSTRACT

The rapid proliferation of Radio Frequency IDentification (RFID) systems realizes integration of physical world with the cyber ones. One of the most promising is the Internet of Things (IoT), a vision in which the Internet extends into our daily activities through wireless networks of uniquely identifiable objects. Given that modern RFID systems are being deployed in large-scale for different applications, without optimizing reader's distribution, many of the readers will be redundant, resulting waste of energy. Additionally, eliminating redundant readers can also decrease probability of reader collisions, as a result, enhancing system performance and efficiency. In this paper, an overlap aware (OA) technique is proposed for eliminating redundant readers. The OA is a distributed approach, which does not need to collect global information for centralizing control, aims to detect maximum amount of redundant readers could be safely removed or turned off with preserving original RFID network coverage. A significant improvement of the OA scheme is that the amount of "write-to-tag" operations could be largely reduced during the redundant reader identification phase. In order to accurately evaluate the performance of the proposed method, it was performed in a variety of scenarios. The experiment results show that the proposed method can provide reliable performance with detecting higher redundancy and has lower algorithm overheads as compared with several well known methods, such as the *RRE*, the *LEO*, the hybrid algorithm (*LEO+RRE*) and the *DRRE*.

## I. INTRODUCTION

Radio Frequency Identifier (RFID) System is an automatic technology aids machines or computers to identify objects, record metadata or control individual target through radio waves. The RFID system is composed by two components, tags and readers. An RFID tag is comprised of integrated circuit with an antenna for storing information and communication, respectively. An RFID reader is capable of reading the information stored at tags located in its sensing range. The electronics in the RFID reader use an outside power resource to generate signal to drives the reader's antenna and turn into radio wave. The radio wave will be received by RFID tag which will reflect the energy in the way of signaling its identification and other related information. In matured RFID systems [1], the reader's RF can also instruct the memory to be read or written from which the tag contained.

Due to its convenience, RFID system is progressively becoming popular in recent years, such as supply chain automation, identification of products at check-out points, security and access control, have been developed to take the primary function of RFID systems. Advantages of RFID technologies, such as price efficiency, fast deployment, reusable and accuracy of stock management also broaden the scope of applications of RFID systems. Advanced characteristics of recent RFID readers, like size miniaturization and capabilities of Wi-Fi [2] or cellular also motivate the development of large-scale RFID systems.

In recent RFID technologies, it is motivated that an RFID system can be integrated with wireless sensor network by interfacing RFID tags with external sensing capabilities, such as light, temperature or shock sensors [3]; forming a hybrid infrastructure [4] that combines advantages of both techniques, such as accurate identification, monitoring of objects and efficient deployment. Similar to wireless sensor network, RFID tags can be deployed in an ad-hoc fashion instead of pre-installed statically. In such way, it will be necessary to install readers in appropriated distance to each other. Otherwise readers would be interfered with each other from the simultaneous operations. The interference could be caused when the frequency band is shared with other potential users. As an RFID reader is designed to accept the tiny signal reflected from a tag. It will be particularly influenced to any relatively powerful transmissions from other readers that happen at the same time. Therefore, efficient methods for detecting redundant readers are of great importance for the development of wireless RFID networks.

While the problem of determining coverage redundancy has been studied in wireless sensor networks, it differs from the redundant RFID reader elimination problem which was proved as NP-hard problem. In this paper, we propose a randomized and de-centralized technique, termed as Overlap Aware optimization (OA), to detect the maximum number of redundant readers that can be safely turned off with preserving the origin network coverage in an RFID network. Advantages of such optimization are twofold; lifetime of wireless RFID network could be extended and reader collisions could be alleviated.

To evaluate performance of the proposed techniques, we have implemented the proposed OA algorithm along with other methods. The experimental results demonstrate that the OA provides superior performance in terms of larger number of redundant reader detected. Both theoretical analysis and performance results show that the OA has lower algorithm overheads, i.e., number of "write-to-tag" operation issued by RFID readers. The performance results also show that the OA is suitable in arbitrary RFID network topology and applicable to large-scale RFID environment in practice.

The rest of this paper is organized as follows: In Section 2, a brief survey of related work will be presented. Section 3 introduces the reader collision problem and redundant reader problem. The proposed redundancy detection algorithm will be introduced in section 4. Performance analysis and simulation comparisons will be given in Section 5. Finally, in Section 6, some concluding remarks are made.

## II. RELATED WORK

In the last decade, research on RFID technologies has been extensively addressed, such as collision problems [5], coverage problems [6], security and privacy problems [7, 8, 9, 10], as well as energy saving.

Standard collision avoidance protocols like RTS-CTS cannot be directly applied in RFID systems due to the reason, in traditional wireless networks, the CTS are sent back to the sender. Similar situation in RFID system, when a reader broadcasts an RTS, all tags in the read range need to send back CTS to the reader [41]. It then requires another collision avoidance mechanism for CTS, and it will make the protocol more complicated. In general, the RFID collision problems can be mainly categorized into two major categories as tag collision [11, 12] and reader collision [13]. Tag collision occurs if multiple tags located in a small area are energized by the reader and then reflect their respective signals back to the reader at the same time. Algorithms based on ALOHA protocol [14, 15, 16, 17, 18, 19, 20] are proposed to avoid tag collision. ALOHA-based algorithm provides the time slot to tags. Tags are allowed to transmit their IDs in the given time slot, and thus the number of probability of tag collision will reduce. Query-tree (*QT*) [21, 22, 23, 24, 25] is another popular protocol for tag anti-collision. In tree-based algorithms, each tag corresponds to the leaf-node of a full binary-tree, and each query string of the reader matches each node of the tree. A tag responds to the reader when the most significant bits of the tag ID match the query string. If responses of tags are collided, the reader will expand the query string and send query again with the new string in next cycles.

The reader collision happens if the interrogation zones of two or more readers overlap. Thus a reader interferes with other reader's operation when both these two readers send query command at the same time. This causes the tag unable to respond the simultaneous queries. The *Colorwave* [26] is a distributed reader anti-collision algorithm based on the TDMA. In this procedure, readers transmit data only in their own colors (timeslots). If the transmission of a reader collides with another reader, the reader will randomly select and reserve a new color. This causes all its neighboring readers to select a new color. Then the reader can avoid collision when all its neighboring readers select a different color. The *Pulse* protocol was referred as beacon broadcast and a CSMA mechanism. Readers periodically in separated control channels send a "beacon" during communication with tags. The contend_back-off and the delay_before_beaconing in the protocol are similar in wireless networks. If the reader receives a beacon, the residual back-off timer will be stored and kept till the next coming chance. This process is expected to achieve the fairness among all readers.

The problem of coverage in wireless sensor networks [27, 28, 40] has been also variety studied. Jiang et al. [29] presented a decentralized and localized density control algorithm that prolongs network lifetime by keeping a minimal number of sensors in active mode while not scarifying any sensing coverage. Tian et al. [30] proposed techniques for detecting redundant sensors whose coverage area is overlapped with others. In addition, Tanaka et al. [31] propose two distributed interference avoidance algorithms based on the detect-and-abort principle for multi-channel readers which can effectively mitigate the reader-to-tag interference as well as the reader-to-reader interference. In [32], Ye et al. presented an energy conserving protocol to extend lifetime of wireless sensor network. The concept of working set was applied in their approach to alternatively turn sensors off and on. A similar research is also presented by Carle et al. [33]. A centralized algorithm was proposed in [34] for organizing sensor network in disjoint subsets of sensors, in order to maximize efficient use of batteries. On the contrary, Zhang et al. [35] proposed a grid based distributed algorithm for maintaining coverage and connectivity. Focusing on RFID system;

Another approach to alleviate reader collision problem is to eliminate redundant readers. A reader is treated as redundant when all the tags in its coverage are also covered by other readers. Carbunar et al. [36] proposed an approximation algorithm, termed as *RRE*, for extending lifetime of wireless RFID reader network. Preserving network coverage and eliminating redundancy in a network, energy efficiency could be improved. Probabilistic analysis and experimental results reported that the proposed heuristic is effective on different network topologies. The *RRE* algorithm was proposed based on a greedy concept for redundant reader elimination. The main idea of the *RRE* is to give priority to the readers which covered most of tags. The reader that covers most number of tags, as compared to other readers in its vicinity, will become the holder of all the tags in its vicinity. Hence the reader that holds no tag will be regarded as redundant. Their extended work [37] has been proposed for addressing both redundancy and coverage detection in sensor network. One of the drawbacks of the RRE algorithm is that each RFID reader needs to write its tag count (number of covered tags) and reader ID on to all its covered tags. This could lead higher transmission overheads and incur higher complexity of write-to-tag operation.

The *LEO* algorithm [38] uses "first come first hold" principle to eliminate redundant readers. A reader that first sends query to a tag will be the holder of the tag. The tag can only be held by a reader once in the *LEO*, so readers that hold no tags will become redundant. The author also proposed a hybrid method to combine the *LEO* with the *RRE* (*LEO+RRE*). The *LEO+RRE* method performs the *LEO* scheme first, and then performs the *RRE* scheme later. The simulation results in showed that the performance of eliminating redundant readers could largely increase if combing the *LEO* with the *RRE*.

The *DRRE* algorithm [39] introduced the concept of the neighboring reader density. In this algorithm, readers that cover the same tags are considered neighbors. The *DRRE* is based on greedy method, too. However, the fundamental of the *DREE* is to give priority to the readers which have more neighbors. Assume that a tag is simultaneously covered by several readers, and then the reader that has the most number of neighbors will be the owner of the tag. Thus the reader that owns no tag is considered redundant. Considering the drawbacks of the present algorithms, in this paper, we propose an efficient redundant reader elimination method, termed as *OA*, which can improve most of the shortcoming of these existing algorithms.

## III. PRELIMINARIES

A reader is redundant if all its covered tags are also covered by at least one of the other readers. Figure 1 shows an RFID network contains three readers, $R_1$-$R_3$, and five tags, $T_1$-$T_5$. Reader $R_2$ is referred as redundant reader because the three tags it covered, i.e., $T_2$, $T_3$ and $T_4$, are also covered by other readers in the same network. Therefore, reader R2 can be safely removed without loss of tags been covered. Advantages of removing redundant readers are twofold; First, because of the limited battery associated with wireless RFID readers, it can extend the lifetime of overall wireless RFID network if the redundant readers are turn off alternatively; Second, the reader to reader interference could be alleviated by eliminating redundant readers. Consequently, reader collisions could be dispelled with the monitoring accuracy of RFID network can be also improved.

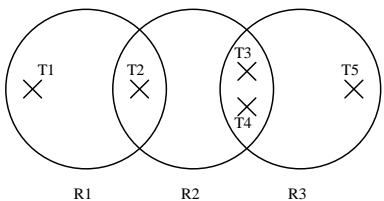

Figure 1: An example of wireless RFID network with redundant reader

A naïve method to detect reader redundancy is to have all readers broadcast a query message to all its covered tags simultaneously. Because RFID tags will reply queries by signaling its id, therefore, if a reader receives no reply, it means that itself a redundant reader. This is either because the reader covers no tag in its covered range, or because tags are not able to reply due to reader collisions.

There are drawbacks of the above method to detect reader redundancy. Firstly, time synchronization among readers is required. Second, network coverage may be destroyed and resulting additional tags uncovered if all redundant readers are turned off. The second situation can be explained by taking the same network topology shown in Figure 1. Let's consider the same readers, $R_1$-$R_3$, but only tags $T_1$-$T_4$ existence in the RFID network. According to the above description, readers $R_2$ and $R_3$ will receive no tag reply and will treat itself as redundant reader. Therefore, if readers $R_2$ and $R_3$ are both turned off, it will result tags $T_3$ and $T_4$ uncovered.

The *RRE* algorithm is a greedy method. The main idea is that a reader, compare with other readers in its vicinity, if the number of covered tags is more than any other readers, then it will be regarded as redundant. To implement this algorithm, the following information should be maintained in tags, *holder* (*H*) and *Tag-Count* (*TC*). The former represents the ID of the reader that covers itself, while the later represents the number of tags been covered by its holder. In the *RRE*, readers will perform the following procedures. Let $S_i$ represents the set of tags that are in reader $R_i$'s vicinity.

- $Count_i$ = the number of tags covered by reader $R_i$.
- For all tags $T_j \in S_i$, if $TC_j < Count_i$, then, $H_j=i$, $TC_j=Count_i$.
- For all tags $T_j \in S_i$, if $H_j \neq i$, then reader $R_i$ is redundant.

<u>Definition 1</u>: The *probability of optimal detection*, denoted as $POD_A$, is the average probability of an algorithm *A*, with which the largest number of redundant readers can be detected by using all possible permutations of reader's execution order; the *probability of redundancy detection*, denoted as $PRD_A$, is the average probability of an algorithm *A*, with which at least one redundant reader can be detected by using all possible permutations of reader's execution order, in a given wireless RFID network.

Given the example shown in Figure 2(a), and the execution order of readers, $R_1 \rightarrow R_2 \rightarrow R_3$, readers $R_1$ and $R_3$ will be finally detected as redundant ones by the *RRE* method, as shown in Figure 3(a). Note that in this example, no matter what the execution order of readers is, the *RRE* method is able to detect the two redundant readers, i.e, $R_1$ and $R_3$. In other words, the *probability of optimal detection* (*POD*) is 100%. However, we can observe that in the RRE method, tags might be recorded with the meta-data multiple times. As a result, the "write-to-tag" operation will be increased directly proportional to the number of readers and tags, say *O*(*NM*), where *N* is the number of tags and *M* is the number of readers..

The *LEO* algorithm aims to reduce the number of write-to-tag operation and increase the lifetime of tags. For this, the *LEO* algorithm only needs one metadata associated with a tag, the *holder* (*H*), showing the ID of the reader that covers it. In *LEO*, readers ($R_i$ as an example) will perform the following procedures.

- For all tags $T_j \in S_i$, if $H_j$ = NULL, then, $H_j=i$.
- If there is no update performed in the last step, then the reader will be regarded as redundant.

Let's use the example in Figure 2(b) to figure out the process of the *LEO* method. We first assume the execution order of readers is, $R_1 \rightarrow R_3 \rightarrow R_2$. The process and results after performing the *LEO* method are given in Figure 3(b), the first half of the notation *A/B*, i.e., *A*, which shows that reader $R_2$ is redundant. As for write-to-tag operation, we can see that tags will be written at most once. If we change the order of reader's execution as $R_1 \rightarrow R_2 \rightarrow R_3$, the results are given in Figure 3(b), the second half of the notation *A/B*, i.e., *B*, showing that no redundant reader can be detected. From this example, we can see that the results of the *LEO* method will be influenced by the execution order of readers though it has a good number of

*write-to-tag* operation, $O(N)$, where $N$ is the number of tags. As for the *probability of optimal detection* (*POD*), in this example, the *LEO* method achieves only 33%. We further look at the results when cross applying the two algorithms in the two examples. For Figure 2(a), we have $PRD_{LEO}$ and $POD_{LEO}$ equal to 66% and 33%, respectively, with a reference to $POD_{RRE}$=100%. For Figure 2(b), both *PRD* and *POD* will be 0% for the *RRE* method.

From the above discussion, we can know that there are pros and cons in the *RRE* and the *LEO* algorithms. An integrated approach, termed as *LEO+RRE* was further proposed to enhance the performance. This hybrid method performs the *LEO* once followed by an additional execution of the *RRE* algorithm. Though it can achieve good *POD*, however, the overhead of *write-to-tag* operation will be a summation of that of the two algorithms, i.e., $O(N)+O(MN)$.

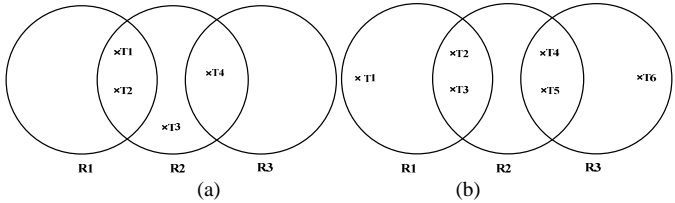

Figure 2: Examples of wireless RFID network with redundant reader

| Reader | $T_1$ | $T_2$ | $T_3$ | $T_4$ |
|---|---|---|---|---|
| $R_{1(2)}$ | $(R_1, 2)$ | $(R_1, 2)$ | | |
| $R_{2(4)}$ | $(R_2, 4)$ | $(R_2, 4)$ | $(R_2, 4)$ | $(R_2, 4)$ |
| $R_{3(1)}$ | | | | |
| Results | $(R_2, 4)$ | $(R_2, 4)$ | $(R_2, 4)$ | $(R_2, 4)$ |

(a)

| Reader | $T_1$ | $T_2$ | $T_3$ | $T_4$ | $T_5$ | $T_6$ |
|---|---|---|---|---|---|---|
| $R_1/R_1$ | $R_1/R_1$ | $R_1/R_1$ | $R_1/R_1$ | | | |
| $R_3/R_2$ | | | | $R_3/R_2$ | $R_3/R_2$ | $R_3/$ |
| $R_2/R_3$ | | | | | | $/R_3$ |
| Results | $R_1/R_1$ | $R_1/R_1$ | $R_1/R_1$ | $R_3/R_2$ | $R_3/R_2$ | $R_3/R_3$ |

(b)

Figure 3: Processes and results of redundant reader identification using different algorithms (a) *RRE* (b) *LEO*

The following statements clarify our network model, research assumptions and characteristics of the proposed *OA* algorithm.

- There is no restriction in the RFID network model. An RFID system could be of arbitrary topology with unlimited number of RFID readers and tags.
- RFID Tags are passive and the associated memory is writable.
- Reader collision problem is assumed avoided before running redundant reader identification.
- The proposed *OA* algorithm is a distributed scheme; which doesn't need to collect global network information for centralizing control. Each reader can perform redundancy detection locally.

IV. THE PROPOSED METHOD

The motivating examples demonstrated in section 3 reveal that the RRE method has higher overheads in writing metadata to tags. In addition, the effectiveness of the *RRE* method mainly relies on tag's distribution, reflecting that the RRE method has worse performance in high density environments. As for the *LEO* algorithm, it has low cost in writing metadata to tags. In addition, the *LEO* has stable performance in either low or high density environments. However, effectiveness of the *LEO* algorithm varies from different execution order of readers. Based on the concept of priority identification, higher overlapped-areas will be analyzed in higher priority. This procedure gives an ideal execution order of readers and increases the detected rate of reader's redundancy. Not only the number of write-to-tag operation is limited, but also the *Probability of Optimal Detection* (*POD*) can be improved. The proposed approach is named as *Overlap Aware Redundant Reader Elimination* (*OA*).

Before introducing the *OA* algorithm, let's look at the example shown in Figure 2(b). According to the explanation in previous section, we have $POD_{RRE}$=0%, $POD_{LEO}$ = $POD_{LEO+RRE}$ = 33%. However, with our humble opinion, it is obvious that reader $R_2$ is redundant. We further observe that Tags $T_2$, $T_3$, $T_4$ and $T_5$ are covered by two readers while tags $T_1$ and $T_6$ are covered by only one reader. So, we classify tags into two types:

Type 1: Tag is covered by exactly one reader
Type 2: Tag is covered by two or more readers

Since there exist type-1 tags in readers $R_1$'s and $R_3$'s vicinity, it is easy to conclude that $R_1$ and $R_3$ are not redundant. On the contrary, if all type-1 tags can be excluded from a reader's vicinity, the reader could be easier to be detected as redundant, as the example of reader $R_2$, shown in Figure 4.

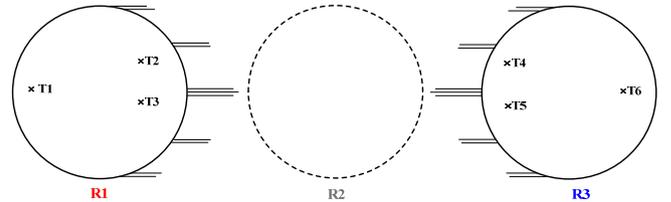

Figure 4: Concept of excluding type-1 tags

To implement the *OA* algorithm, the following information should be maintained in tags: *holder* (*H*) and *Status* (*ST*), the former is the ID of the reader that covers the tag; the later indicates status of the tag, *overlap* or *lock*. In the *OA* method, readers will perform the following procedures to identify its redundancy.

- For all tags $T_j \in S_i$, if $H_j$ = NULL, then $H_j=i$.
- For all tags $T_j \in S_i$, if ($H_j \neq i$ && $ST_j$=NULL) then $ST_j$=*overlap*.
- For all tags $T_j \in S_i$, if ($H_j \neq i$ && $ST_j \neq$NULL) then $ST_j$=*lock*.
- For all tags $T_j \in S_i$, if there exist a tag $T_j$, s.t. $P(i, j)$ is true, then reader $R_i$ is not redundant; otherwise, $R_i$ is redundant, where $P(i,j)=(H_j=i$ && $ST_j \neq lock)$.

From the above procedures we can see that, in the *OA* algorithm, the amount of *write-to-tag* operation is irrelative to the number of readers. In worst case, each step could lead at most one "write" operation, as a result, the complexity of *write-to-tag* operation is $O(N)$, showing the same degree as that of the *LEO* method.

Let's use the same example in Figure 2(b) to clarify the process of the *OA* algorithm. We first assume the execution order of readers is, $R_2 \rightarrow R_1 \rightarrow R_3$. The detailed processes with performing the *OA* method are given in Figure 5. In step 1, readers write its ID to the tags in its vicinity as shown in Figure 5(a). In step 2, readers $R_1$ and $R_3$ write *overlap* information to the tags with different holder in its vicinity, as shown in Figure 5(b). In step 3, readers $R_1$ and $R_3$ write *luck* information to the tags with different holder in its vicinity if the tag's status is not null, as shown in Figure 5(c). In step 4, reader $R_2$ is aware its redundancy, and turn it off.

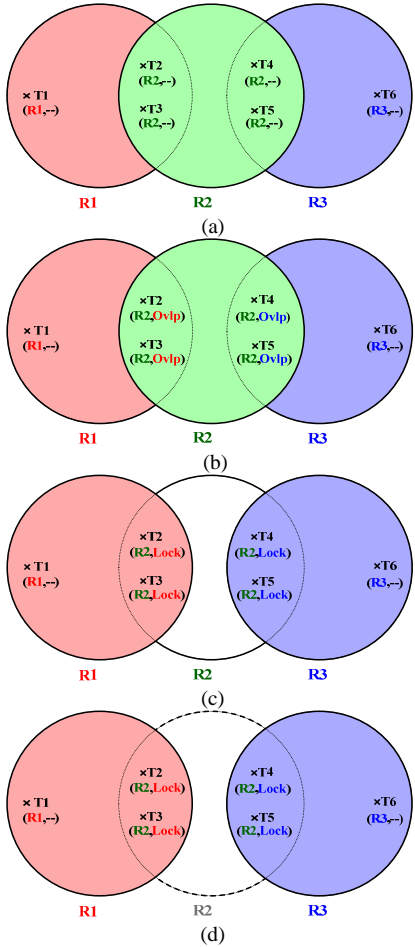

Figure 5: Processes of redundant reader detection using *OA*. Example I (a) step 1 (b) step 2 (c) step 3 (d) step 4

The above example shows that the *OA* algorithm is able to detect $R_2$ as a redundant reader under the execution order, $R_2 \rightarrow R_1 \rightarrow R_3$. Since there are six permutations for three readers, Figure 6 outlines the results of redundant reader detection under the six different reader's execution order. The *OA* algorithm is reported in the rightmost column. In all the six cases, the *OA* is able to detect the redundant reader. Analysis for the *LEO*, *RRE* and *LEO+RRE* is also summarized. The *Probability of Optimal Detection* (*POD*) of these algorithms can be obtained accordingly, as given in the end row.

To better understand the effectiveness of the *OA* algorithm in different network topology, let's use the graph shown in Figure 2(a) as our second example. Given the execution order of $R_1 \rightarrow R_2 \rightarrow R_3$, the detailed process of the *OA* method is depicted in Figure 7. In step 1, readers write its ID to the tags in its vicinity as shown in Figure 7(a). In step 2, reader $R_2$ writes *overlap* information to the tags with different holder in its vicinity, as shown in Figure 7(b). In step 3, reader $R_2$ writes *luck* information to the tags with different holder in its vicinity if the tag's status is not null, as shown in Figure 7(c). In step 4, readers $R_1$ and $R_3$ are aware its redundancy. Figure 8 summarizes the *Probability of Optimal Detection* (*POD*) of the *OA* and the compared algorithms.

| Execution order | LEO | RRE | LEO+RRE | OA |
|---|---|---|---|---|
| $R_1 \rightarrow R_2 \rightarrow R_3$ | | | | $R_2$ |
| $R_1 \rightarrow R_3 \rightarrow R_2$ | $R_2$ | | $R_2$ | $R_2$ |
| $R_2 \rightarrow R_1 \rightarrow R_3$ | | | | $R_2$ |
| $R_2 \rightarrow R_3 \rightarrow R_1$ | | | | $R_2$ |
| $R_3 \rightarrow R_1 \rightarrow R_2$ | $R_2$ | | $R_2$ | $R_2$ |
| $R_3 \rightarrow R_2 \rightarrow R_1$ | | | | $R_2$ |
| Probability (*POD*) | 33% | 0% | 33% | 100% |

Figure 6: Effectiveness under different reader's execution order

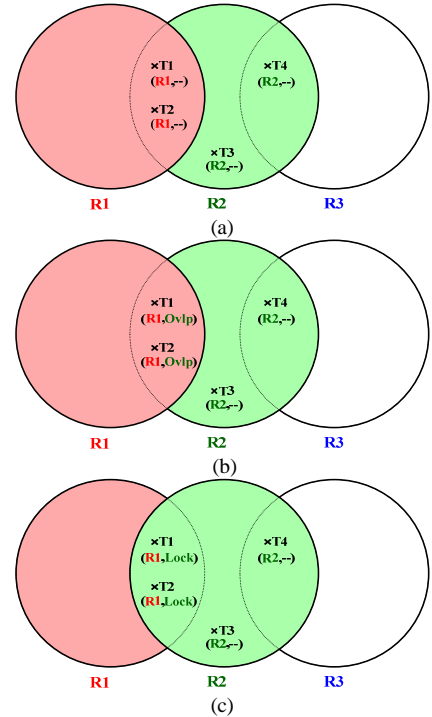

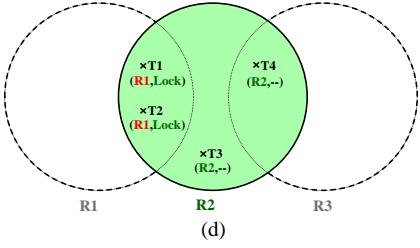

Figure 7: Processes of redundant reader detection using *OA*. Example II (a) step 1 (b) step 2 (c) step 3 (d) step 4

|  | *LEO* | *RRE* | *LEO+RRE* | *OA* |
|---|---|---|---|---|
| Probability (*POD*) | 33% | 100% | 100% | 100% |

Figure 8: Effectiveness under different reader's execution order

Further analysis was conducted based on the 3rd and 4th examples shown in Figure 9. Example III shows the situation that all tags in the network are covered by at least two readers. Example IV simulates high density environments, in which readers $R_1$, $R_3$, and $R_4$ are neighbored by at least two adjacent readers. Statistics of *Probability of Optimal Detection* (*POD*) of the *OA* and the compared algorithms, for examples III and IV, are summarized in Figure 10.

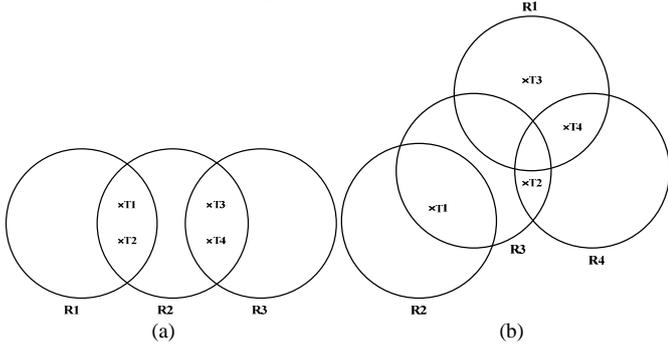

Figure 9: Additional examples of wireless RFID network with redundant reader (a) example III (b) example IV

|  | *LEO* | *RRE* | *LEO+RRE* | *OA* |
|---|---|---|---|---|
| Probability (*POD*) | 33.3% | 100% | 66.6% | 66.6% |

(a)

|  | *LEO* | *RRE* | *LEO+RRE* | *OA* |
|---|---|---|---|---|
| Probability (*POD*) | 20.8% | 29.1% | 33.3% | 62.5% |

(b)

Figure 10: Statistics of *POD* (a) Example III (b) example IV

The four examples illustrated in this section represent different characteristics of redundancy. According to the comprehensive analysis, we can see that the *OA* is reliable in detecting redundant readers. As for the write-to-tag operation, the *OA* algorithm has the same complexity as of the *LEO* method, $O(N)$, which is much less than the *RRE*'s $O(NM)$.

## V. PERFORMANCE ANALYSIS

To evaluate the performance of the proposed method, we have implemented the OA algorithm, along with other previous ones, such as *RRE*, *LEO*, *LRRE* (*LEO+RRE*), *DRRE* and *SDRRE*. All programs were written in MinGW C++ under the Code Block developing environment. In the simulated environment, tags and readers are randomly deployed in a 10000×10000 square, an open area with no obstacles, and readers have the same radius (500 units) of the interrogation zone. In order to simulate the effectiveness of these algorithms in different topologies of RFID network, four scenarios were conducted in our experiments. Parameters of experiment setup for the scenario setup are showed in Table 1.

TABLE I. PARAMETERS OF EXPERIMENT SETUP

| Setup | Number of Readers (*NR*) | Number of tags (*NT*) | Radius (*R*ad) |
|---|---|---|---|
| I (low density) | 500 | 100-1000 | 500 |
| II (high density) | 500 | 1000-10000 | 500 |
| III (vary *NR*) | 100-500 | 10000 | 500 |
| IV (vary *R*ad) |  |  |  |

Figure 11(a) compares the amount of redundant reader detected by these algorithms in scenario I. We can see that in very low density (*NT*=100) situation, most algorithms have similar results. The *OA* algorithm outperforms other algorithms when *NT* increases to 300. The *LEO* approach performs worst in this scenario, and the performance of the *RRE* is obviously affected by the total number of tags. Comparison on the cost of *write-to-tag* operation is given in Figure 11(b). The *DRRE* and *SDRRE* obviously have the highest amount of *write-to-tag* operation, which is regarded as impracticable for implementation. Though the *LEO* has the lowest cost in I/O, the detection rate still needs to be improved. Note that the cost of *write-to-tag* operation in *OA* is only a bit higher than the *LEO*.

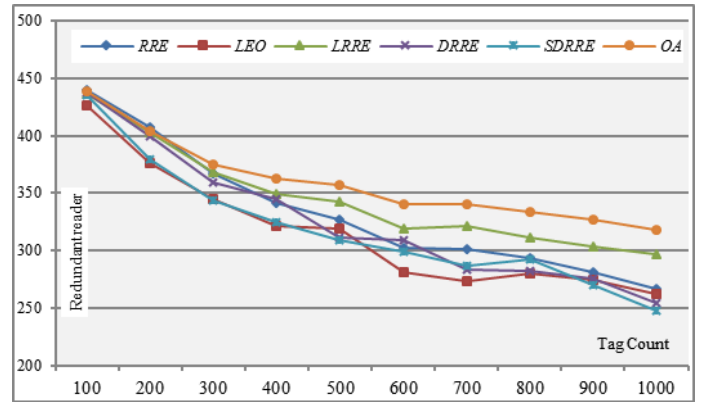

(a)

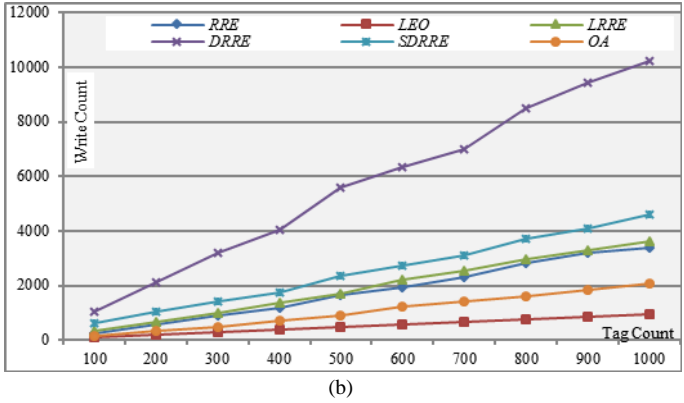

Figure 11: Performance Comparison of setup I (a) redundant reader detection (b) *write-to-tag* operation

Figure 12 gives the amount of redundant reader detected by these algorithms in scenario II. In high density network, he *OA* algorithm keeps its superiority in all test cases. The *LEO* method outperforms the *RRE* when *NR* increases to 2000. This implies that the *RRE* method is not suitable in high density environments. As for the *write-to-tag* operation, it is expected that it will be increased linearly as the number of tags increased. From Figure 12(b), we can see that the simulation results in this test are similar to that of the scenario I.

significant in these algorithms. The *OA* continues its superior performance in these tests. As for write-to-tag operation, the cost of *OA* and *LEO* are irrelative to the number of readers and remain in a low cost. For those *RRE*-like algorithms, such as *RRE* and *DRRE*, the write count increases significantly when number of reader becomes large.

Figure 14 presents the performance of these algorithms in scenario IV, which varies the radius of readers sensing range. The *OA* algorithm still performs best. As for the *write count*, because the increase of reader's sensing radius, the amount of tags been covered by two or more readers will be increased as well. As a result, the *OA* needs a bit more *write* operation to the tags deployed in those overlapped areas, thus, worse than the *LEO* approach. Despite of this, the *OA* still has lower I/O cost as compare to all the other approaches, such as *RRE*, *LRRE*, *DRRE* and *SDRRE*. The following are remarks from the above experimental results.

- The *RRE* algorithm is suitable only in low density RFID environments
- The *LEO* algorithm performs neutral but has the lowest cost on the "*write-to-tag*" operation.
- The *DRRE* algorithm performs neutral but has the highest cost for the "*write-to-tag*" operation.
- The *OA* algorithm outperforms all other compared approaches in both low- and high-density circumstances, in terms of redundant reader detection.

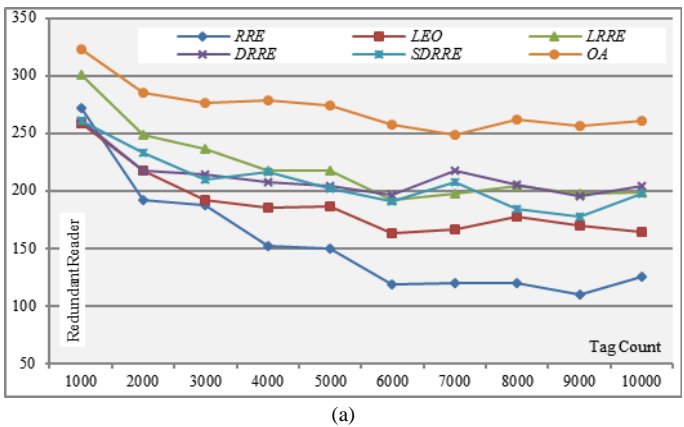

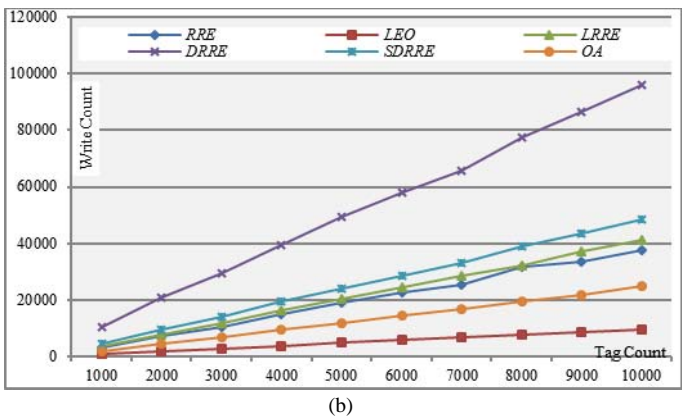

Figure 12: Performance Comparison of setup II (a) redundant reader detection (b) *write-to-tag* operation

Figure 13 accesses the performance of these algorithms in scenario III. We can see that the effect of *reader count* is not

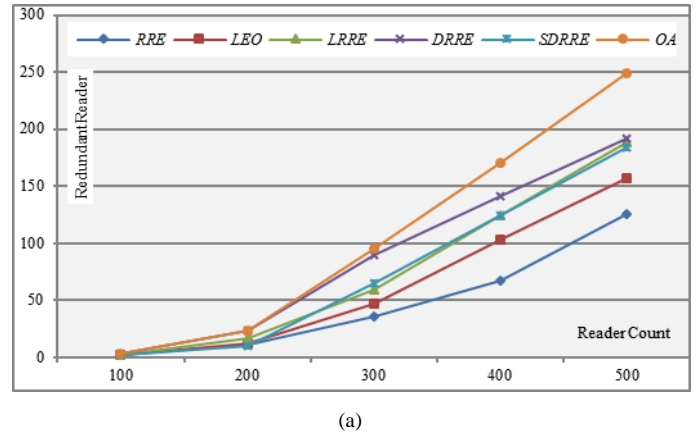

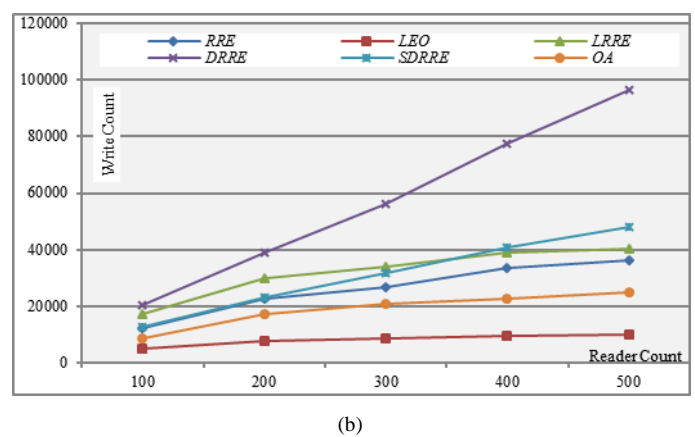

Figure 13: Performance Comparison of setup III (a) redundant reader detection (b) *write-to-tag* operation

## VI. CONCLUSIONS AND FUTURE WORK

In this paper, we presented an *overlap-aware* (*OA*) technique for detecting redundant readers in wireless RFID networks. The proposed algorithm is a distributed approach, which does not need to collect global information for centralizing control, aims to detect maximum amount of redundant readers could be safely removed or turned off with preserving original RFID network coverage. A significant improvement of the OA scheme is that the amount of "write-to-tag" operations could be largely reduced during the redundant reader identification phase. In order to evaluate the performance of the proposed method, the proposed algorithm was performed in three different scenarios, low density, high density and hybrid environments. Our extensive simulations show that the proposed algorithm is very accurate and efficient when compared to several well known methods, such as *RRE*, *LEO*, *LEO+RRE* and *DRRE*, with an enhanced ratio of detected redundancy by 86.24%, 46%, 24% and 28%, respectively. As for future work, establishing theoretical models to assess the impact of redundant reader elimination and reader/tag I/O optimization on energy saving is of interest to be investigated.

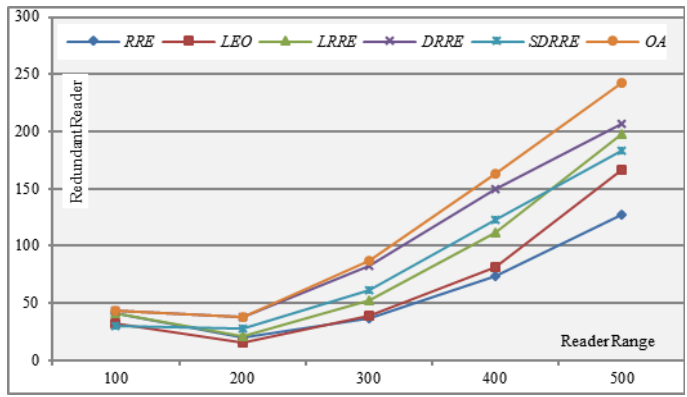

(a)

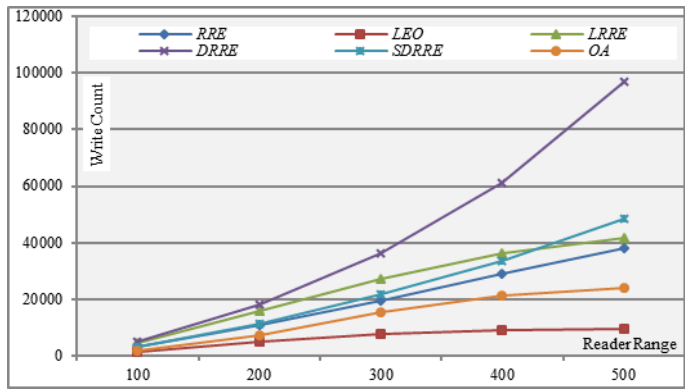

(b)

Figure 14: Performance Comparison of setup IV (a) redundant reader detection (b) *write-to-tag* operation

## ACKNOWLEDGEMENTS

This work is supported by the National Natural Science Foundation of China (Grant No. 61103185), Natural Science Foundation of the Higher Education Institutions of Jiangsu Province, China (Grant No. 11KJB520009), and the 9th Six Talents Peak Project of Jiangsu Province (Grant No. DZXX-043).